\documentclass{eptcs-modified}
 \providecommand{\event}{} 
 \usepackage{breakurl}             
\usepackage{amsmath}
\usepackage{amssymb}
\usepackage{amsthm}
\usepackage{graphics}

\long\def\greybox#1{%
    \newbox\contentbox%
    \newbox\bkgdbox%
    \setbox\contentbox\hbox to \hsize{%
        \vtop{
            \kern\columnsep
            \hbox to \hsize{%
                \kern\columnsep%
                \advance\hsize by -2\columnsep%
                \setlength{\textwidth}{\hsize}%
                \vbox{
                    \parskip=\baselineskip
                    \parindent=0bp
                    #1
                }%
                \kern\columnsep%
            }%
            \kern\columnsep%
        }%
    }%
    \setbox\bkgdbox\vbox{
        \pdfliteral{0.85 0.85 0.85 rg}
        \hrule width  \wd\contentbox %
               height \ht\contentbox %
               depth  \dp\contentbox
        \pdfliteral{0 0 0 rg}
    }%
    \wd\bkgdbox=0bp%
    \vbox{\hbox to \hsize{\box\bkgdbox\box\contentbox}}%
    \vskip\baselineskip%
}

\title{Towards Synthetic Descriptive Set Theory: An instantiation with represented spaces}
\author{
Arno Pauly
\institute{Clare College\\ University of Cambridge, United Kingdom}
\email{Arno.Pauly@cl.cam.ac.uk}
\and
Matthew de Brecht
\institute{National Institute of Information and Communications
Technology\\Kyoto, Japan\email{matthew@nict.go.jp}}
}
\def\titlerunning{Synthetic Descriptive Set Theory}
\def\authorrunning{A. Pauly \& M. de Brecht}

\begin{document}
\theoremstyle{definition}
\newtheorem{theorem}{Theorem}
\newtheorem{definition}[theorem]{Definition}
\newtheorem{problem}[theorem]{Problem}
\newtheorem{assumption}[theorem]{Assumption}
\newtheorem{corollary}[theorem]{Corollary}
\newtheorem{proposition}[theorem]{Proposition}
\newtheorem{lemma}[theorem]{Lemma}
\newtheorem{observation}[theorem]{Observation}
\newtheorem{fact}[theorem]{Fact}
\newtheorem{question}[theorem]{Open Question}
\newtheorem{conjecture}[theorem]{Conjecture}
\newcommand{\dom}{\operatorname{dom}}
\newcommand{\id}{\textnormal{id}}
\newcommand{\Cantor}{{\{0, 1\}^\mathbb{N}}}
\newcommand{\Baire}{{\mathbb{N}^\mathbb{N}}}
\newcommand{\Lev}{\textnormal{Lev}}
\newcommand{\hide}[1]{}
\newcommand{\mto}{\rightrightarrows}
\newcommand{\uint}{{[0, 1]}}
\newcommand{\bft}{\mathrm{BFT}}
\newcommand{\lbft}{\textnormal{Linear-}\mathrm{BFT}}
\newcommand{\pbft}{\textnormal{Poly-}\mathrm{BFT}}
\newcommand{\sbft}{\textnormal{Smooth-}\mathrm{BFT}}
\newcommand{\ivt}{\mathrm{IVT}}
\newcommand{\cc}{\textrm{CC}}
\newcommand{\lpo}{\textrm{LPO}}
\newcommand{\llpo}{\textrm{LLPO}}
\newcommand{\aou}{AoU}
\newcommand{\Ctwo}{C_{\{0, 1\}}}
\newcommand{\name}[1]{\textsc{#1}}
\newcommand{\C}{\textrm{C}}
\newcommand{\ic}[1]{\textrm{C}_{\sharp #1}}
\newcommand{\xc}[1]{\textrm{XC}_{#1}}
\newcommand{\me}{\name{P}.~}
\newcommand{\etal}{et al.~}
\newcommand{\eval}{\operatorname{eval}}
\newcommand{\Sierp}{Sierpi\'nski }
\newcommand{\isempty}{\operatorname{IsEmpty}}

\newcounter{saveenumi}
\newcommand{\seti}{\setcounter{saveenumi}{\value{enumi}}}
\newcommand{\conti}{\setcounter{enumi}{\value{saveenumi}}}

\maketitle

\begin{abstract}
Using ideas from synthetic topology, a new approach to descriptive set theory is suggested. Synthetic descriptive set theory promises elegant explanations for various phenomena in both classic and effective descriptive set theory. Presently, we mainly focus on developing the ideas in the category of represented spaces.
\end{abstract}

\section{Introduction}
\emph{Synthetic descriptive set theory} is the idea that descriptive set theory can be reinterpreted as the study of certain endofunctors and derived concepts, primarily in the category of represented spaces. It is proposed as an abstract framework explaining the many similarities between descriptive set theory (e.g.~\cite{kechris}), effective descriptive set theory (e.g.~\cite{moschovakis}) and facets of recursion theory. A crucial novel aspect is that classes of functions such as the $\Sigma_n^0$-measurable ones that are commonly seen as generalizations of the continuous functions are now considered to be a special case -- thus making the observation that they share many properties with the continuous functions trivial. The change in viewpoint proceeds via the recognition that the concepts $\Sigma_n^0$, $\Delta_n^0$, $\Delta_1^1$, etc., can be considered as endofunctors acting on a suitable category.

This research programme can be seen as a continuation of \name{Escard\'o}'s \emph{synthetic topology} \cite{escardo}. While the central concepts can be formulated in a generic setting of a cartesian closed category with a \Sierp-space-like object, we focus on the expression of these general concepts in the category of represented spaces that underlies the TTE-approach to computable analysis (\cite{weihrauchd}). This category does seem to be a very appropriate setting for descriptive set theory, in particular it contains the structures considered in classical descriptive set theory such as separable metric spaces or Borel equivalence relations.

To some extent we can view this work as reinterpreting the classical field of descriptive set theory
into a kind of type theory, where spaces are types and the endofunctors are
certain kinds of modal operators. There is a long history of these kinds of
type theories in theoretical computer science, such as \name{Moggi}'s work
\cite{moggi} which models computational semantics
with monads, and also various flavors of Jean-Yves \name{Girard}'s \emph{Light Linear
Logic} \cite{girard2} which use substructural logic and modal operators to characterize
computational complexity. In synthetic descriptive set theory, the modal
operators characterize topological complexity or non-constructiveness.

A development inside descriptive set theory that to some extent mirrors the introduction of notions derived from computing machines suggested here is the use of games to characterize function classes. Pioneered by \name{Wadge} \cite{wadge, wadge2}, a culmination can be found in \cite{motto-ros3} by \name{Motto-Ros}. \name{Nobrega} has provided a translation of these results into the language of Weihrauch degrees in his Master's thesis \cite{nobrega}, which makes them even more accessible for our purposes.

Various results in the literature can -- in hindsight -- be read as contributing to synthetic descriptive set theory, this pertains to \cite{brattka,debrecht4,moschovakis2, paulydebrecht,kihara3, kihara3b, debrecht5,selivanov4,selivanov4b} by \name{Brattka}, \name{Moschovakis}, \name{Higuchi}, \name{Kihara}, \name{Schr\"oder}, \name{Selivanov} and the authors (and this list quite certainly is incomplete).

After recalling (extremely briefly) some of the relevant concepts from synthetic topology on the one hand, and then from descriptive set theory on the other hand, we will present the new ideas in two main sections. First, the core ideas of \emph{synthetic} descriptive theory are introduced, in a primarily category-theoretic language. The reader may find it difficult to see the connections to classical descriptive set theory until the next section. There, concrete endofunctors are examined regarding their connections to well-known concepts in descriptive set theory. The proofs in Section \ref{section:coreconcepts} are generally extremely short; while Section \ref{sec:examples} would see some longer (though not very involved) proofs. The latter can be seen as reflecting some less-elegant aspects of traditional definitions in descriptive set theory.

\subsection{Synthetic topology}
The core idea of synthetic topology is that in any cartesian closed category (i.e.~a category allowing the formation of function spaces) which has a special object $\mathbb{S}$ behaving suitably like the \Sierp-space in topology, it is possible to introduce a variety of concepts from topology, such as the space of open (closed, compact) subsets of a given space, and properties of spaces such as being compact or being Hausdorff. In this the morphisms of the category are pretended to be the continuous functions. \cite{escardo,escardo2,bauer4} introduced and developed synthetic topology; while \name{Taylor}'s \emph{abstract stone duality} \cite{taylor, taylor2} features some similar ideas.

\emph{Admissibility} as a property of objects in a category subjected to synthetic topology can, following the work of \name{Schr\"oder} \cite{schroder, schroder5}, be understood as marking those space whose behaviour in the category (as codomain of morphisms) is fully determined by their topological properties. In essence, the admissible spaces in a cartesian closed category form a cartesian closed subcategory that also is a subcategory\footnote{Note however that the admissibly represented spaces do not inherit their product from $Top$, but rather from the category of sequential spaces. In other words, the topology on the product of two admissibly represented spaces is the sequentialization of the product topology}. of the category $Top$ of topological spaces and continuous maps.

A self-contained treatment of synthetic topology instanced with the category of represented spaces can be found in \cite{pauly-synthetic-arxiv}. Here, we just recall a few formal definitions relevant for the development of synthetic descriptive set theory:

A represented space is a pair $(X, \delta_X) := \mathbf{X}$ where $\delta_X : \subseteq \Cantor \to X$ is a partial surjection. For $f : \mathbf{X} \to \mathbf{Y}$ and $F : \subseteq \Cantor \to \Cantor$, we call $F$ a realizer of $f$ (notation $F \vdash f$), iff $\delta_Y(F(p)) = f(\delta_X(p))$ for all $p \in \dom(f\delta_X)$. A map between represented spaces is called computable (continuous), iff it has a computable (continuous) realizer.
A priori, the notion of a continuous map between represented spaces and a continuous map between topological spaces are distinct and should not be confused!

We consider two categories of represented spaces, one equipped with the computable maps, and one equipped with the continuous maps. We call the resulting structure a \emph{category extension} (cf.~\cite{paulysearchproblems}), as the former is a subcategory of the latter, and shares its structure (products, coproducts, exponentials).

The set of continuous functions from $\mathbf{X}$ to $\mathbf{Y}$ can be turned into a represented space $\mathcal{C}(\mathbf{X}, \mathbf{Y})$ itself, with the evaluation map being computable due to the UTM-theorem. This establishes our category to be cartesian closed.

We want to make use of two special represented spaces, $\mathbb{N} = (\mathbb{N}, \delta_\mathbb{N})$ and $\mathbb{S} = (\{\bot, \top\}, \delta_\mathbb{S})$. The representation are given by $\delta_\mathbb{N}(0^n10^\mathbb{N}) = n$, $\delta_\mathbb{S}(0^\mathbb{N}) = \bot$ and $\delta_\mathbb{S}(p) = \top$ for $p \neq 0^\mathbb{N}$. Computability on $\mathbb{N}$ coincides with the classical notion of computability. The functions $\wedge, \vee : \mathbb{S} \times \mathbb{S} \to \mathbb{S}$ and $\bigvee : \mathcal{C}(\mathbb{N}, \mathbb{S}) \to \mathbb{S}$ are computable.

Now we define the set of open subsets of a space $\mathbf{X}$ to be the set of functions $\mathcal{C}(\mathbf{X}, \mathbb{S})$, where we identify a set with its characteristic function. We immediately obtain that $(f, U) \mapsto f^{-1}(U) : \mathcal{C}(\mathbf{X}, \mathbf{Y}) \times \mathcal{O}(\mathbf{Y}) \to \mathcal{O}(\mathbf{X})$ is computable for all represented spaces $\mathbf{X}$, $\mathbf{Y}$ -- this is just composition of functions! Hence, any continuous function between represented spaces matches the definition of continuity for functions between topological spaces. An alternative formulation is that $f \mapsto f^{-1} : \mathcal{C}(\mathbf{X}, \mathbf{Y}) \to \mathcal{C}(\mathcal{O}(\mathbf{Y}), \mathcal{O}(\mathbf{X}))$ is computable.

Given a represented space $\mathbf{X}$, consider the map $\kappa_\mathbf{X} : \mathbf{X} \to \mathcal{O}(\mathcal{O}(\mathbf{X}))$ mapping any point to its neighborhood filter, i.e.~$\kappa(x) = \{U \in \mathcal{O}(\mathbf{X}) \mid x \in U\}$. Let $\mathbf{X}_\kappa$ be the image of $\kappa_\mathbf{X}$. Now we call $\mathbf{X}$ (computably) admissible, if $\kappa$ is injective and $\kappa^{-1} : \mathbf{X}_\kappa \to \mathbf{X}$ is continuous (computable). Note that $\mathbf{X}_\kappa$ is always computably admissible, i.e.~isomorphic to $(\mathbf{X}_\kappa)_\kappa$.

Now a space $\mathbf{Y}$ is (computable) admissible if and only if the map $f \mapsto f^{-1} : \mathcal{C}(\mathbf{X}, \mathbf{Y}) \to \mathcal{C}(\mathcal{O}(\mathbf{Y}), \mathcal{O}(\mathbf{X}))$ is continuously (computably) invertible. Hence, for admissible spaces, the inherent (represented space) definition of continuity coincides with the topological version.

\subsection{Descriptive Set Theory}
A central part of descriptive set theory is the Borel hierarchy. Consider a separable metric space $\mathbf{X}$. Now let $\Sigma_1^0(\mathbf{X}) := \mathcal{O}(\mathbf{X})$, $\Pi_\alpha^0(\mathbf{X}) := \{X \setminus U \mid U \in \Sigma_\alpha^0(\mathbf{X})\}$, $\Sigma_{\alpha+1}^0(\mathbf{X}) = \{\bigcup_{i \in \mathbb{N}} A_i \mid \forall i \in \mathbb{N} \ A_i \in \Pi_\alpha^0(\mathbf{X})\}$ and $\Sigma_{\beta}^0(\mathbf{X}) = \bigcup_{\alpha < \beta} \Sigma_{\alpha}^0(\mathbf{X})$ for limit ordinals $\beta$. Moreover, let $\Delta_\alpha^0(\mathbf{X}) = \Sigma_\alpha^0(\mathbf{X}) \cap \Pi_\alpha^0(\mathbf{X})$.

The $\Sigma_\alpha^0$-sets behave in some ways like the open set: They are closed under countable unions and finite intersections, and the preimages of a $\Sigma_\alpha^0$-set under a continuous function is a $\Sigma_\alpha^0$-set again. We also find that $\Sigma_\alpha^0(\mathbf{X}) \subseteq \Sigma_{\alpha'}^0(\mathbf{X})$ if $\alpha < \alpha'$.

For non-metric topological spaces that are still countably based and $T_0$, \name{Selivanov} \cite{selivanov3} suggest a modified definition of the Borel hierarchy, using $\Sigma_{\alpha+1}^0(\mathbf{X}) := \{\bigcup_{i \in \mathbb{N}} (U_i \setminus U_i') \mid \forall i \in \mathbb{N} \ U_i, U'_i \in \Sigma_\alpha^0(\mathbf{X})\}$ instead. This modification ensures that $\Sigma_\alpha^0(\mathbf{X}) \subseteq \Sigma_{\alpha'}^0(\mathbf{X})$ if $\alpha < \alpha'$ remains true, and is equivalent to the original definitions for metric spaces.

If we start only with the effectively open sets, and demand all countable unions to be uniform, we obtain the effective Borel hierarchy instead. Formalizing the uniformity conditions for the countable unions can be slightly cumbersome, and is omitted here.

Let $\mathfrak{B} \in \{\Sigma_\alpha^0, \Pi_\alpha^0, \Delta_\alpha^0\}$. We call a function $f : \mathbf{X} \to \mathbf{Y}$ $\mathfrak{B}$-measurable, if $f^{-1}(U) \in \mathfrak{B}(\mathbf{X})$ for any $U \in \mathcal{O}(\mathbf{Y})$. A common theme in descriptive set theory is to provide alternative characterizations of some class of $\mathfrak{B}$-measurable functions.

Say that the Baire class 0 functions are the continuous functions, the Baire class $1$ functions the $\Sigma_2^0$-measurable functions\footnote{For some special metric spaces, the following theorem would hold without explicitly demanding truth for $\alpha = 1$, i.e.~with the Baire class 1 functions being the point-wise limits of continuous functions. This fails for other spaces, though: If $\mathbf{X}$ is connected and $\mathbf{Y}$ discrete, then point-wise limits of continuous functions are continuous themselves (Example taken from \cite{mottoros4}). Such exceptions marring the theory disappear when moving to the synthetic approach.}, the Baire class $\alpha$ functions the point-wise limits functions of Baire class $< \alpha$. Now we can formulate the:

\begin{theorem}[Lebesgue -- Hausdorff -- Banach]
\label{theo:lhbt}
Let $\mathbf{X}$, $\mathbf{Y}$ be separable metric spaces. Then a function $f : \mathbf{X} \to \mathbf{Y}$ is Baire class $\alpha$ iff it is $\Sigma_{\alpha+1}^0$-measurable.
\end{theorem}

Next, we shall call a function $f : \mathbf{X} \to \mathbf{Y}$ piecewise continuous, if there is a cover $(A_i)_{i \in \mathbb{N}}$ of $\mathbf{X}$ of closed sets (i.e.~$\Pi_1^0$-sets), such that any $f_{|A_i}$ is continuous. The corresponding characterization result is:
\begin{theorem}[Jayne \& Rogers \cite{jaynerogers}\footnote{See also the simplified proof in \cite{ros,ros2}.}]
Let $\mathbf{X}$ be Polish and $\mathbf{Y}$ be separable. Then a function $f : \mathbf{X} \to \mathbf{Y}$ is piecewise continuous iff it is $\Delta_2^0$-measurable.
\end{theorem}

\section{Core concepts of synthetic descriptive set theory}
\label{section:coreconcepts}
Our investigation starts with endofunctors on the category of continuous functions between represented spaces. An endofunctor is an operation $d$ from a category to itself which maps objects to objects, morphisms to morphisms, preserves identity morphisms,
and is compatible with composition, i.e.~$d(f \circ g) = (df) \circ (dg)$. For any two represented spaces $\mathbf{X}$, $\mathbf{Y}$, an endofunctor $d$ induces\footnote{In the presence of exponentials and a final object, an endofunctor $d$ may have an internal characterization. For fixed objects $\mathbf{X}$, $\mathbf{Y}$, let $D : \mathcal{C}(\mathbf{X}, \mathbf{Y}) \to \mathcal{C}(d\mathbf{X}, d\mathbf{Y})$ be an internal realization of $d$, if the following holds: Let $f : \mathbf{X} \to \mathbf{Y}$ be a morphism, and $f' : \mathbf{X} \times \mathbf{1} \to \mathbf{Y}$ the corresponding morphism up to equivalence. By definition of the exponential, we then have a map $\lambda f' : \mathbf{1} \to \mathcal{C}(\mathbf{X}, \mathbf{Y})$. In the same way, there is a map $\lambda (df)' : \mathbf{1} \to \mathcal{C}(d\mathbf{X}, d\mathbf{Y})$. The criterion now is $\lambda (df)' = D \circ \lambda f'$.} a map $d : \mathcal{C}(\mathbf{X}, \mathbf{Y}) \to \mathcal{C}(d\mathbf{X}, d\mathbf{Y})$. If this map is always computable, we call $d$ computable\footnote{Modulo the continuity/computability distinction, this would be a special case of an enriched endofunctor, if we understand a cartesian closed category to be enriched over itself.}.  In the following, $d$ shall always be some computable endofunctor.

The typical examples relevant for our development of descriptive set theory will be operators that keep the underlying set of a represented spaces the same, and modify the representation in a sufficiently uniform way to ensure the requirements for computable endofunctors. Such operators have been called \emph{jump operators} in \cite{debrecht5}, and specific examples can be found both there and in Section \ref{sec:examples}. For computable endofunctors that do change the underlying sets in a significant way, the interpretation of many of the following definitions becomes less clear, but an example of a computable endofunctor that still produces sensible notions is given in Subsection \ref{subsec:compacts}.

The computable endofunctors we study correspond to classes of sets such as $\Sigma_2^0$, $\Sigma_3^0$, $\Delta_2^0$, etc.; with the closure properties of the set-classes depending on how the endofunctor interacts with products. We say that $d$ preserves binary products, if $d(\mathbf{X} \times \mathbf{X}) \cong d\mathbf{X} \times d\mathbf{X}$ (where $\cong$ denotes computable isomorphism) for any represented space $\mathbf{X}$, and that $d$ preserves products if $d\mathcal{C}(\mathbf{N}, \mathbf{X}) \cong \mathcal{C}(\mathbb{N}, d\mathbf{X})$ for any represented space $\mathbf{X}$.

\subsection{The $d$-open sets}
For a represented space $\mathbf{X}$, we shall call $\mathcal{C}(\mathbf{X}, d\mathbb{S})$ the space of $d$-open sets $\mathcal{O}^d(\mathbf{X})$. If $d\mathbf{S}$ still has the underlying set $\{\bot, \top\}$ the elements of $\mathcal{O}^d(\mathbf{X})$ actually are subsets of $X$ in the usual way\footnote{Which is to identify a function $f : X \to \{\bot, \top\}$ with the set $f^{-1}(\top)$, and vice versa a set with its characteristic function.}, otherwise this is a purely abstract definition\footnote{Predicates in fuzzy logic would be an example of an entity analogue to the characteristic function of a set that crucially has not $\{\bot, \top\}$ as codomain. A somewhat similar example is presented in Subsection \ref{subsec:compacts}.}. The complements of $d$-open sets are $d$-closed sets, denoted by $\mathcal{A}^d(\mathbf{X})$. A variety of nice closure properties follows immediately, with the proofs being straight-forward modifications of those for the corresponding results for open sets in \cite[Proposition 6]{pauly-synthetic-arxiv}:

\begin{proposition}
\label{prop:dopensets}
The following operations are computable for any represented spaces $\mathbf{X}, \mathbf{Y}$:
\begin{enumerate}
\item $(f, U) \mapsto f^{-1}(U) : \mathcal{C}(\mathbf{X}, \mathbf{Y}) \times \mathcal{O}^d(\mathbf{Y}) \to \mathcal{O}^d(\mathbf{X})$
\item $\operatorname{Cut} : \mathbf{Y} \times \mathcal{O}^d(\mathbf{X} \times \mathbf{Y}) \to \mathcal{O}^d(\mathbf{X})$ mapping $(y, U)$ to $\{x \mid (x, y) \in U\}$
    \seti
\end{enumerate}
If $d$ preserves binary products, we additionally obtain:
\begin{enumerate}
\conti
\item $\cap, \cup : \mathcal{O}^d(\mathbf{X}) \times \mathcal{O}^d(\mathbf{X}) \to \mathcal{O}^d(\mathbf{X})$
\item $\times : \mathcal{O}^d(\mathbf{X}) \times \mathcal{O}^d(\mathbf{Y}) \to \mathcal{O}^d(\mathbf{X} \times \mathbf{Y})$
\seti
\end{enumerate}
If $d$ preserves products, we additionally obtain:
\begin{enumerate}
\conti
\item $\bigcup : \mathcal{C}(\mathbb{N}, \mathcal{O}^d(\mathbf{X})) \to \mathcal{O}^d(\mathbf{X})$
\end{enumerate}
\begin{proof}
\begin{enumerate}
\item This is just function composition.
\item And this is partial evaluation.
\item Given that $\wedge, \vee : \mathbb{S} \times \mathbb{S} \to \mathbb{S}$ are computable functions, and $d$ is a computable endofunctor, we find that $\wedge, \vee : d(\mathbb{S} \times \mathbb{S}) \to d\mathbb{S}$ are computable. Now $d$ is assumed to preserve binary products, and $\cap, \cup$ are obtained by composing with $\wedge, \vee$.
\item This uses again computable $\wedge : d\mathbb{S} \times d\mathbb{S} \to d\mathbb{S}$, together with type-conversion.
\item If $d$ is a computable endofunctor preserving products, we can obtain computable $\bigcup : \mathcal{C}(\mathbb{N}, d\mathbb{S}) \to d\mathbb{S}$ from computable $\bigcup : \mathcal{C}(\mathbb{N}, \mathbb{S}) \to \mathbb{S}$. The rest is function composition.
\end{enumerate}
\end{proof}
\end{proposition}

\subsection{$d$-continuity and $d$-measurability}
Now we can introduce the notion of $d$-measurability: We call a function $f : \mathbf{X} \to \mathbf{Y}$ $d$-measurable, if $f^{-1} : \mathcal{O}(\mathbf{Y}) \to \mathcal{O}^d(\mathbf{X})$ is well-defined and continuous, i.e.~if the preimages of open sets under $f$ are uniformly $d$-open. The $d$-measurable functions from $\mathbf{X}$ to $\mathbf{Y}$ thus form a represented space $\mathcal{C}^d(\mathbf{X}, \mathbf{Y})$, which is by construction homeomorphic to a subspace of $\mathcal{C}(\mathcal{O}(\mathbf{Y}), \mathcal{O}^d(\mathbf{X}))$.

A $d$-continuous function from $\mathbf{X}$ to $\mathbf{Y}$ shall just be a continuous function $f : \mathbf{X} \to d\mathbf{Y}$. Note again, that if $d$ alters the underlying sets, then a $d$-continuous function between represented spaces will not necessarily induce a function on the underlying sets. The notion of $d$-continuity is a generalization of the Kleisli-morphisms w.r.t.~a monad -- if $d$ can be turned into a monad, then the $d$-continuous functions are precisely the Kleisli-morphisms. Some, but not all, of our examples of computable endofunctors will actually be monads in a natural way.

Reminiscent of the Banach-Lebesgue-Hausdorff theorem and the Jayne Rogers theorem, a $d$-measurable function is characterized by how preimages of open sets behave (similar to the $\mathfrak{B}$-measurable function), while $d$-continuous functions are characterized by what information is available on their function values (similar to Baire class $\alpha$ functions or piecewise continuous functions).

Both the $d$-measurable and the $d$-continuous functions have some of the closure properties expected from classes of $\mathfrak{B}$-measurable functions. For their formulation, note that the composition of two computable endofunctors is a computable endofunctor again.

\begin{proposition}
\label{prop:dcontdmeas}
Both $d$-continuous and $d$-measurable maps are closed under composition with continuous maps from both sides, i.e.~the following maps are computable for any represented spaces $\mathbf{X}$, $\mathbf{Y}$, $\mathbf{Z}$:
\begin{enumerate}
\item $\circ : \mathcal{C}(\mathbf{X}, \mathbf{Y}) \times \mathcal{C}(\mathbf{Y}, d\mathbf{Z}) \to \mathcal{C}(\mathbf{X}, d\mathbf{Z})$
\item $\circ : \mathcal{C}(\mathbf{X}, \mathbf{Y}) \times \mathcal{C}^d(\mathbf{Y}, \mathbf{Z}) \to \mathcal{C}^d(\mathbf{X}, \mathbf{Z})$
\item $\circ : \mathcal{C}(\mathbf{X}, d\mathbf{Y}) \times \mathcal{C}(\mathbf{Y}, \mathbf{Z}) \to \mathcal{C}(\mathbf{X}, d\mathbf{Z})$
\item $\circ : \mathcal{C}^d(\mathbf{X}, \mathbf{Y}) \times \mathcal{C}(\mathbf{Y}, \mathbf{Z}) \to \mathcal{C}^d(\mathbf{X}, \mathbf{Z})$
\seti
\end{enumerate}
More generally, we can consider a second computable endofunctor $e$ and obtain:
\begin{enumerate}
\conti
\item $\circ : \mathcal{C}(\mathbf{X}, e\mathbf{Y}) \times \mathcal{C}(\mathbf{Y}, d\mathbf{Z}) \to \mathcal{C}(\mathbf{X}, ed\mathbf{Z})$
\seti
\end{enumerate}
Taking into consideration the definition of $\mathcal{O}^d(\mathbf{X})$ as $\mathcal{C}(\mathbf{X}, d\mathbb{S})$, we get the special case:
\begin{enumerate}
\conti
\item $(f, U) : \mathcal{C}(\mathbf{X}, e\mathbf{Y}) \times \mathcal{O}^d(\mathbf{Y}) \to \mathcal{O}^{ed}(\mathbf{X})$
\item $\circ : \mathcal{C}(\mathbf{X}, e\mathbf{Y}) \times \mathcal{C}^d(\mathbf{Y}, \mathbf{Z}) \to \mathcal{C}^{ed}(\mathbf{X}, \mathbf{Z})$
\seti
\end{enumerate}
Finally, we find that $e$-continuity uniformly implies $e$-measurability:
\begin{enumerate}
\conti
\item $\id : \mathcal{C}(\mathbf{X}, e\mathbf{Y}) \to \mathcal{C}^e(\mathbf{X}, \mathbf{Y})$
\end{enumerate}
\begin{proof}
\begin{enumerate}
\item Just regular function composition.
\item Consider $\mathcal{C}^d(\mathbf{Y}, \mathbf{Z})$ as (homeomorphic to) a subspace of $\mathcal{C}(\mathcal{O}(\mathbf{Z}), \mathcal{O}^d(\mathbf{Y}))$, likewise for $\mathcal{C}^d(\mathbf{X}, \mathbf{Z})$. Recall that $\mathcal{O}^d(\mathbf{Y})$ is essentially $\mathcal{C}(\mathbf{Y}, d\mathbb{S})$. Now $(f, g) \mapsto (h \mapsto (f \circ g(h)))$ realizes the desired functional.
\item As $d$ is a computable endofunctor, we can move from $\mathcal{C}(\mathbf{Y}, \mathbf{Z})$ as second argument to $\mathcal{C}(d\mathbf{Y}, d\mathbf{Z})$, and then use regular function composition.
\item Let us view $\mathcal{C}^d(\mathbf{X}, \mathbf{Y})$ and $\mathcal{C}^d(\mathbf{X}, \mathbf{Z})$ as (homeomorphic to) subspaces of $\mathcal{C}(\mathcal{O}(\mathbf{Y}), \mathcal{O}^d(\mathbf{Z}))$ and $\mathcal{C}(\mathcal{O}(\mathbf{Z}), \mathcal{O}^d(\mathbf{X}))$ respectively. Now $(f, g) \mapsto (h \mapsto f(g \circ h))$ realizes the desired functional.
\item As $e$ is a computable endofunctor, we can move from $\mathcal{C}(\mathbf{Y}, d\mathbf{Z})$ as second argument to $\mathcal{C}(e\mathbf{Y}, ed\mathbf{Z})$, and then use regular function composition.
\item Choose $\mathbf{Z} := \mathbb{S}$ in (5.).
\item Type conversion together with (6.).
\item By currying and considering $d := \id$ in (6.).
\end{enumerate}
\end{proof}
\end{proposition}

\subsection{$d$-admissibility}
Having seen that $d$-continuity always implies $d$-measurability, we now strive for conditions that make the converse implication true, as well. Noting that $\id$-continuity is continuity of maps between represented spaces, and $\id$-measurability (uniform) topological continuity, we see that we need a notion of $d$-admissibility.

As a special case of Proposition \ref{prop:dcontdmeas} (8) with $\mathbf{X} = \mathbf{1}$ and using trivial isomorphisms, we obtain the computability of a canonic mapping $\kappa^d : d\mathbf{Y} \to \mathcal{C}(\mathcal{O}(\mathbf{Y}), d\mathbb{S})$. The image of $d\mathbf{Y}$ under $\kappa^d$ shall be denoted by $\kappa^d\mathbf{Y}$ (not by $\kappa^dd\mathbf{Y}$!!).

\begin{proposition}
\label{prop-admissibility}
The following are equivalent:
\begin{enumerate}
\item $\id : \mathcal{C}(\mathbf{X}, d\mathbf{Y}) \to \mathcal{C}^d(\mathbf{X}, \mathbf{Y})$ is computably invertible for any represented space $\mathbf{X}$.
\item $\kappa^d : d\mathbf{Y} \to \mathcal{C}(\mathcal{O}(\mathbf{Y}), d\mathbb{S})$ is computably invertible.
\item $\kappa^d\mathbf{Y} \cong d\mathbf{Y}$.
\end{enumerate}
\begin{proof}
\begin{description}
\item[$1. \Rightarrow 2.$] Choose $\mathbf{X} := \mathrm{1}$, and use the canonic isomorphism $\mathcal{C}(\mathrm{1}, \mathbf{Z}) \cong \mathbf{Z}$ twice.
\item[$2. \Rightarrow 3.$] The map $\kappa^d : d\mathbf{Y} \to \mathcal{C}(\mathcal{O}(\mathbf{Y}), d\mathbb{S})$ is always computable: Given $y \in d\mathbf{Y}$ and $U \in \mathcal{O}(\mathbf{Y}) = \mathcal{C}(\mathbf{Y}, \mathbf{S})$, we start by moving to $dU \in \mathcal{C}(d\mathbf{Y}, d\mathbf{S})$ using that $d$ is a computable endofunctor. Then we apply $dU$ to $y$; what remains is currying. That $\kappa^d$ has a computable inverse is asserted as $(2.)$, the claim then follows directly.
\item[$3. \Rightarrow 1.$] Under the assumption $\kappa^d\mathbf{Y} \cong d\mathbf{Y}$, we may show instead that $\id : \mathcal{C}(\mathbf{X}, \kappa^d\mathbf{Y}) \to \mathcal{C}^d(\mathbf{X}, \mathbf{Y})$ is computable invertible. Given $f^{-1} \in \mathcal{C}^d(\mathbf{X}, \mathbf{Y}) \subseteq \mathcal{C}(\mathcal{O}(\mathbf{Y}), \mathcal{C}(\mathbf{X}, d\mathbb{S}))$ and $x \in \mathbf{X}$, we may obtain $(U \mapsto f^{-1}(U)(x)) : \mathcal{O}(\mathbf{Y}) \to d\mathbf{S}$. This in turn is $\kappa^d(f(x)) \in \kappa^d\mathbf{Y}$, so again just currying remains to be done.
\end{description}
\end{proof}
\end{proposition}

A space $\mathbf{Y}$ satisfying these equivalent conditions shall be called $d$-admissible. We observe the following:

\begin{proposition}
\label{sadmissible}
$\mathbb{S}$ is $d$-admissible.
\begin{proof}
We need to show that $\kappa^d : d\mathbb{S} \to \mathcal{C}(\mathcal{O}(\mathbb{S}), d\mathbb{S})$ is computably invertible. To do this, simply substitute $\{\top\} \in \mathcal{O}(\mathbb{S})$ at the corresponding position.
\end{proof}
\end{proposition}

Now, consider $\kappa^d$ as an operation on the whole category of continuous functions between represented spaces. It is not hard to verify that $\kappa^d$ itself is a computable endofunctor. Even more, we can consider $d \mapsto \kappa^d$ ($=: \kappa$) as an operation on computable endofunctors! As a consequence of Proposition \ref{sadmissible} we obtain:

\begin{corollary}
$\kappa^{(\kappa^d)} \cong \kappa^d$.
\begin{proof}
Note that the right hand side of $\kappa^d : d\mathbf{Y} \to \mathcal{C}(\mathcal{O}(\mathbf{Y}), d\mathbb{S})$  depends on $d$ only via $d\mathbb{S}$. So any $\kappa^{(\kappa^d)} : \kappa^d\mathbf{X} \to \kappa^{(\kappa^d)}\mathbf{X}$ essentially is the identify. Hence, $d\mathbb{S} = \kappa^d\mathbb{S}$ from Proposition \ref{sadmissible} yields the claim.
\end{proof}
\end{corollary}

\begin{corollary}
Every represented space is $\kappa^d$-admissible.
\end{corollary}

\begin{corollary}
$\mathcal{O}^d(\mathbf{X}) = \mathcal{O}^{\kappa^d}(\mathbf{X})$.
\end{corollary}

\begin{corollary}
$d$-measurability and $\kappa^d$-continuity coincide.
\end{corollary}

\begin{corollary}
If $\mathbf{Y}$ is $d$-admissible, then $\mathcal{C}^d(\mathbf{X}, \mathbf{Y})$ and $\mathcal{C}(\mathbf{X}, d\mathbf{Y})$ are homeomorphic.
\end{corollary}

\begin{corollary}
If $\mathbf{Y}$ is $e$-admissible, then $\circ : \mathcal{C}^e(\mathbf{X}, \mathbf{Y}) \times \mathcal{C}^d(\mathbf{Y}, \mathbf{Z}) \to \mathcal{C}^{ed}(\mathbf{X}, \mathbf{Z})$ is computable.
\end{corollary}

For a large class of spaces and computable endofunctors, we can provide admissibility results without having to resort to modifying the endofunctor. We start with the seemingly innocuous:
\begin{proposition}
Let $d$ preserve products. Then $\mathcal{O}(\mathbb{N})$ is $d$-admissible.
\begin{proof}
By assumption, $d\mathcal{O}(\mathbb{N}) \cong \mathcal{C}(\mathbb{N}, d\mathbb{S}) \cong \mathcal{O}^d(\mathbb{N})$. Now consider the right hand side of Proposition \ref{prop-admissibility} (2). As $\mathbb{N}$ is admissible, we find that we may go from the induced subspace of $\mathcal{C}(\mathcal{O}(\mathcal{O}(\mathbb{N})), d\mathbb{S})$ back to $\mathcal{C}(\mathbb{N}, d\mathbb{S})$, thus obtaining the desired equivalence.
\end{proof}
\end{proposition}

\begin{corollary}
Let $d$ preserve products, and let $\mathbf{X}$ be countably based and admissible. Then $\mathbf{X}$ is $d$-admissible.
\end{corollary}

The preceding corollary relies on \name{Weihrauch}'s observation \cite{weihrauchd} that the countably-based admissible spaces are just the subspaces of $\mathcal{O}(\mathbb{N})$, together with $d$-admissibility being closed under formation of subspaces. Additionally, it may be the reason that countably-based $T_0$-spaces seem to form a natural demarkation line for the extension of descriptive set theory \cite{debrecht6}. Combining its statement with Proposition \ref{prop:dopensets}, we see that any computable endofunctor preserving products nicely characterizes a $\Sigma$-like class of sets and the corresponding measurable functions on all countably based admissible spaces.

\subsection{Further concepts}
The other concepts from synthetic topology studied for represented spaces in \cite{pauly-synthetic-arxiv}, namely Hausdorff, discreteness, compactness and overtness, can also be lifted along some endofunctor, and retain most of their nice properties. Rather than listing all of these statements and definitions, we shall only consider those used later in applications.

\begin{definition}
A space $\mathbf{X}$ is called computably $d$-Hausdorff, iff $x \mapsto \{x\} : \mathbf{X} \to \mathcal{A}^d(\mathbf{X})$ is computable.
\end{definition}

\begin{proposition}
The following are equivalent:
\begin{enumerate}
\item $\mathbf{X}$ is computably $d$-Hausdorff.
\item $\mathalpha{\neq} : \mathbf{X} \times \mathbf{X} \to d\mathbb{S}$ is computable.
\seti
\end{enumerate}
If $d$ preserves binary products, then the following are also equivalent to those above:
\begin{enumerate}
\conti
\item $\{(x,x) \mid x \in \mathbf{X}\} \in \mathcal{A}^d(\mathbf{X} \times \mathbf{X})$ is computable.
\item $\operatorname{Graph} : \mathcal{C}(\mathbf{Y}, \mathbf{X}) \to \mathcal{A}^d(\mathbf{Y} \times \mathbf{X})$ is well-defined and computable for any represented space $\mathbf{Y}$.
\end{enumerate}
\end{proposition}

Some properties related to $d$-Hausdorff have been studied by \name{Schr\"oder} and \name{Selivanov} in \cite{selivanov4,selivanov4b}.

\begin{proposition}
If $d$ preserves binary products and $\mathbf{X}$ is computably Hausdorff, then $d\mathbf{X}$ is computably $d$-Hausdorff.
\end{proposition}

\begin{definition}
A space $\mathbf{X}$ is called $d$-overt, iff $\operatorname{IsNonEmpty} : \mathcal{O}^d(\mathbf{X}) \to d\mathbb{S}$ is computable.
\end{definition}

\begin{proposition}
If $f : \mathbf{X} \to \mathbf{Y}$ is a computable surjection and $\mathbf{X}$ is $d$-overt, then so is $\mathbf{Y}$.
\end{proposition}

\begin{proposition}
\label{prop:overtunion}
If $d$ preserves products and each $\mathbf{X}_n$ is $d$-overt, then so is $\bigcup_{n \in \mathbb{N}} \mathbf{X}_n$.
\end{proposition}

\hide{
\begin{proposition}
$\operatorname{Lift} : \mathcal{C}(\mathbf{X},\mathbf{Y}) \to \mathcal{C}(\mathcal{O}^d(\mathcal{O}^d(\mathbf{X})), \mathcal{O}^d(\mathcal{O}^d(\mathbf{Y})))$ defined by $U \in \operatorname{Lift}(f)(V)$ iff $f^{-1}(U) \in V$ is computable.
\end{proposition}

The preceding observation can be strengthened in case that $dd \cong d$ (thus in particular if $d$ is a monad):
\begin{proposition}
Let $dd \cong d$. Then $\operatorname{Lift} : \mathcal{C}(\mathbf{X},d\mathbf{Y}) \to \mathcal{C}(\mathcal{O}^d(\mathcal{O}^d(\mathbf{X})), \mathcal{O}^d(\mathcal{O}^d(\mathbf{Y})))$ defined by $U \in \operatorname{Lift}(f)(V)$ iff $f^{-1}(U) \in V$ is computable.
\end{proposition}}

\subsection{The Markov-variant}
In effective descriptive set theory, the notion of affectivity between higher-order objects being employed often is not computability, but rather Markov computability. A function $f : \mathbf{X} \to \mathbf{Y}$ is called Markov-computable, if there is some computable partial  function $\phi : \subseteq \mathbb{N} \to \mathbb{N}$, such that whenever $i$ is an index of a computable element in $\mathbf{X}$, then $\phi(i)$ is an index of $f(i)$. Any computable function is Markov-computable, while the converse fails.

Subsequently, an endofunctor $d$ is called Markov-computable, if any $d : \mathcal{C}(\mathbf{X}, \mathbf{Y}) \to \mathcal{C}(d\mathbf{X}, d\mathbf{Y})$ is Markov-computable\footnote{Note that just as a computable endofunctor is linked to the topological jump operators of \cite{debrecht5}, Markov-computable endofunctors are linked to the computable jump operators.}. The effective measurability notion going with Markov-computable endofunctors is (weak) non-uniform computability, i.e.~if for any computable $U \in \mathcal{O}(\mathbf{Y})$ we find $f^{-1}(U) \in \mathcal{O}^d(\mathbf{X})$ to be computable, we call $f$ to be Markov-$d$-measurable. The represented space $\mathcal{C}^{\mathcal{M}d}(\mathbf{X}, \mathbf{Y})$ of Markov-$d$-measurable functions essentially represents a function by some oracle $p$ paired with a table listing indices of computably open sets and their $p$-computably $d$-open preimages.

\begin{proposition}
Let $d$ be Markov-computable. Then $\id : \mathcal{C}(\mathbf{X}, d\mathbf{Y}) \to \mathcal{C}^{\mathcal{M}d}(\mathbf{X}, \mathbf{Y})$ is computable.
\begin{proof}
Let us be given a function $f \in \mathcal{C}(\mathbf{X}, d\mathbf{Y})$, i.e. we have an index $n$ and an oracle $p$ that realize $f$. The oracle involved is retained. To construct the table, let us further be given an index $i$ for a computable $U \in \mathcal{O}(\mathbf{Y})$. By $d$ being Markov-computable, we can obtain an index $j$ for $dU : d\mathbf{Y} \to d\mathbf{S}$. Composing the machines of $n$ and $j$ yields an index for $f^{-1}(U)$ relative to $p$.
\end{proof}
\end{proposition}

We can define the Markov-variant of $\kappa^d$ via letting $\eta^d : d\mathbf{Y} \to \mathcal{C}^{\mathcal{M}d}(\mathbf{1}, \mathbf{Y})$ be the canonic map, and subsequently obtain a notion of Markov-$d$-admissibility with just the same properties as before.

\subsection{Adjoint endofunctors}
\label{subsec:adjointness}
A computable endofunctor $d$ is computably-left-adjoint to a computable endofunctor $e$ (and $e$ is right-adjoint to $d$), if $\mathcal{C}(d\mathbf{X},\mathbf{Y})$ and $\mathcal{C}(\mathbf{X},e\mathbf{Y})$ are computably isomorphic, and the isomorphisms are natural in $\mathbf{X}$ and $\mathbf{Y}$.

Likewise, a Markov-computable endofunctor $d$ is Markov-computably-left-adjoint to a Markov-computable endofunctor $d$, if $\mathcal{C}(d\mathbf{X},\mathbf{Y})$ and $\mathcal{C}(\mathbf{X},e\mathbf{Y})$ are Markov-computably isomorphic, and the isomorphisms are natural in $\mathbf{X}$ and $\mathbf{Y}$. Note that a statement that $\mathbf{X}$ and $\mathbf{Y}$ are Markov-computably isomorphic only refers to the cardinality (as there has to be a bijection) and to the computable elements. Note further that for computable endofunctors being Markov-computably-adjoint is a weaker condition than being computably adjoint (and that both concepts formally make sense).

At the current state, we do not have interesting examples of pairs of computably-adjoint computable endofunctors. We will discuss two cases of Markov-computably adjoint Markov-computable functors later.

It is quite illuminating to see the special case of the definitions above where $\mathbf{Y} := \mathbb{S}$. We see that if $d$ is (Markov)-computably-left-adjoint to $e$, then the (computably) $e$-open subsets of $\mathbf{X}$ are precisely the (computably) open subsets of $d\mathbf{X}$. This aspect of our two examples below has been utilized before.

It is a central fact in the study of pairs of adjoint functors in category theory that their composition induces a monad. Some consequences of this of interest for our theory are the following:
\begin{proposition}
\label{prop:adjoint}
Let the (Markov)-computable endofunctor $d$ be (Markov)-computably-left-adjoint to the (Markov)-computable endofunctor $e$. Then:
\begin{enumerate}
\item There is a canonic computable\footnote{$\eta_\mathbf{X}$ is indeed always computable, even if the endofunctors involved are only Markov-computable. The same pattern applies in the following.} unit map $\eta_\mathbf{X} : \mathbf{X} \to ed\mathbf{X}$.
\item $e\mathbf{Y}$ and $ede\mathbf{Y}$ are computably isomorphic.
\item $\mathalpha{\circ} : \mathcal{C}(\mathbf{X},ed\mathbf{Y}) \times \mathcal{C}(\mathbf{Y},e\mathbf{Z}) \to \mathcal{C}(\mathbf{X},e\mathbf{Z})$ is well-defined and computable.
\item $(de) \cong (de)(de)$.
\end{enumerate}
\begin{proof}
\begin{enumerate}
\item $\eta_\mathbf{X} \in \mathcal{C}(\mathbf{X},ed\mathbf{X})$ is the image of the computable map $\id_{d\mathbf{X}} \in \mathcal{C}(d\mathbf{X},d\mathbf{X})$ under the assumed (Markov)-computable isomorphism, and Markov-computable isomorphisms map computable points to computable points.
\item $\eta_{e\mathbf{Y}} : e\mathbf{Y} \to ede\mathbf{Y}$ from $(1.)$ provides one direction. The computable inverse is obtained by starting from computable $\id_{e\mathbf{Y}} \in \mathcal{C}(e\mathbf{Y},e\mathbf{Y})$, moving to the corresponding map under the computable isomorphism in $\mathcal{C}(de\mathbf{Y},e\mathbf{Y})$ and then applying the endofunctor $e$ on both sides to reach $\id : ede\mathbf{Y} \to e\mathbf{Y}$.
\item By combining $(2.)$ with Proposition \ref{prop:dcontdmeas} $(5.)$.
\item A direct consequence of $(2.)$.
\end{enumerate}
\end{proof}
\end{proposition}

Items $(3.) \& (4.)$ in the preceding proposition shows that if, given some endofunctor $e$, we can find a Markov-computably-left-adjoint $d$ for it, then we obtain a class functions (namely the $de$-continuous ones) that is closed under composition, and that if composed with an $e$-continuous function from the right, again yield an $e$-continuous function.

The importance of adjointness had already been noticed in \cite{debrecht5}.

\section{Examples}
\label{sec:examples}
To substantiate our claim that the framework of $d$-admissible spaces actually pertains to descriptive set theory, a few computable endofunctors are investigated. These endofunctors are not freshly introduced here, but have been studied for a while, in particular in work by \name{Ziegler} \cite{ziegler2, ziegler3}. As we cannot (yet?) give generic characterizations of these endofunctors in terms not specific for the category of represented spaces, we do leave behind our proto-synthetic framework at this stage.

\subsection{$\Sigma_\alpha^0$-measurability}
Consider the partial function $\lim : \subseteq \Baire \to \Baire$ defined via $\lim(p)(n) = \lim_{i \to \infty} p(\langle n, i\rangle)$. This induces a computable endofunctor $'$ via $(X, \delta_X)' = (X, \delta_X \circ \lim)$ and $(f : \mathbf{X} \to \mathbf{Y})' = f : \mathbf{X}' \to \mathbf{Y}'$. We iterate this endofunctor, so let $\mathbf{X}^{(0)} = \mathbf{X}$, $\mathbf{X}^{(\alpha+1)} = (\mathbf{X}^{(\alpha)})'$ and $\mathbf{X}^{(\beta)} = \pi_2(\coprod_{\gamma < \beta} \mathbf{X}^{(\gamma)})$ for limit\footnote{The definition for limit ordinals was suggested by \name{Bauer} at CCA 2009.} ordinals $\beta$. We claim that the $^{(\alpha)}$-open subsets of a represented space are a suitable generalization of the $\Sigma_{\alpha+1}^0$-subsets of a metric space.

\begin{proposition}
$^{(\alpha)}$ is a computable endofunctor preserving binary products. Moreover, $^{(\alpha+1)}$ even preserves products.
\begin{proof}
That $^{(\alpha)}$ is an endofunctor follows directly from its definition, which leaves the underlying sets and set-theoretical functions unchanged. It being computable for limit ordinal $\alpha$ is straight-forward by slice-wise application, so it suffices to prove that $'$ is computable. Let $F \vdash f : \mathbf{X} \to \mathbf{Y}$. A naive attempt to obtain a realizer $F'$ for $f' : \mathbf{X}' \to \mathbf{Y}'$ would be to define $F'(\langle p_0, p_1, \ldots\rangle) = \langle F(p_0), F(p_1), \ldots\rangle$. However, any $F(p_i)$ may fail to be well-defined as an element of $\Baire$. The algorithm will, however, have to produce initial segments of the output of increasing length if $\lim_{i \in \mathbb{N}} p_i \in \dom(F)$. So instead, let $F^n$ be the modification of the algorithm for $F$ that runs only for time $n$, and then stops. Now let $\lambda_{n,i} = \max \{j \leq n \mid F^n(p_j)(i) \textnormal{ exists}\}$, and $F'(\langle p_0, p_1, \ldots\rangle)(\langle n, i\rangle) = F(p_{\lambda_{n,i}})(i)$.

That $^{(\alpha+1)}$ preserves products is a consequence of the position-wise definition of convergence for sequences. To see that $^{(\alpha)}$ preserves binary products for limit ordinals $\alpha$, we just need that $\sup \{\beta_1, \beta_2\} < \alpha$ for $\beta_1, \beta_2 < \alpha$ (and the failure of this to generalize to countably many $\beta$'s is the reason why for limit ordinal $\alpha$, $^{(\alpha)}$ will not preserve countable products).
\end{proof}
\end{proposition}

\begin{corollary}[Synthetic Lebesgue -- Hausdorff -- Banach Theorem]
\label{corr:syntlhbt}
Every countably based admissible space is $^{(\alpha+1)}$-admissible.
\end{corollary}

Most of the closure properties of the $^{(\alpha)}$-open subsets follow directly from the general case. Additionally, as $\id : \mathbf{X} \to \mathbf{X}'$ and $\neg : \mathbb{S} \to \mathbb{S}'$ are computable, we obtain that $\id, ^C: \mathcal{O}^{(\alpha)}(\mathbf{X}) \to \mathcal{O}^{(\alpha+1)}(\mathbf{X})$ are computable\footnote{Generally, $\id : \mathcal{O}^{(\alpha)}(\mathbf{X}) \to \mathcal{O}^{(\beta)}(\mathbf{X})$ is continuous for $\beta \geq \alpha$, and computable, if $\beta$ is computable relative to $\alpha$.}. By combining these results, also $\bigcup(\cdot^C) : \mathcal{C}(\mathbb{N}, \mathcal{O}^{(\alpha)}(\mathbf{X})) \to \mathcal{O}^{(\alpha+1)}(\mathbf{X})$ becomes computable.

\begin{proposition}
\label{prop:sigma2decomposition}
Let $\mathbf{X}$ be a computable metric space. Then $\bigcup(\cdot^C) : \mathcal{C}(\mathbb{N}, \mathcal{O}(\mathbf{X})) \to \mathcal{O}^{'}(\mathbf{X})$ admits a computable multi-valued inverse.
\begin{proof}
First, we show the claim for $\mathbf{X} = \Cantor$, then we transfer the result to general computable metric spaces using the fact that those have effectively fiber-compact representations.

Assume we have some realizer $\chi$ of some $U \in \mathcal{O}'(\Cantor)$. As $\delta_{\mathbf{S}} \circ \lim(q) = \top$ iff $\exists n \ \lim_{i \to \infty} q(\langle n, i\rangle) = 1$ iff $\exists n \exists k \forall i \geq k q(\langle n, i\rangle) = 1$, we have that $p \in U$ iff $\exists n \exists k \forall i \geq k \chi(p)(\langle n, i\rangle) = 1$. Now consider the closed sets $A_{n,k} = \{p \mid \forall i \geq k \ \chi(p)(\langle n, i\rangle) = 1\}$ and notice $U = \bigcup_{n,k \in \mathbb{N}} A_{n,k}$ and that the $A_{n,k}$ can by construction be computed from $\chi$.

Before proceeding to general computable metric spaces, we point out that the preceding proof carries over rather directly to subspaces of $\Cantor$, totality of the maps involved is not a concern.

Now, given some $U \in \mathcal{O}'(\mathbf{X})$, we use Proposition \ref{prop:dcontdmeas} (6) to compute $\delta_\mathbf{X}^{-1}(U) \in \mathcal{O}'(\dom (\delta_\mathbf{X}))$ and use the established result to obtain some $(A_i)_{i \in \mathbb{N}}$ with $A_i \in \mathcal{A}(\Cantor)$ and $\delta_\mathbf{X} \left (\bigcup_{i \in \mathbb{N}} A_i\right) = \bigcup_{i \in \mathbb{N}} \delta_\mathbf{X}(A_i) = U$. Now notice that for $A \in \mathcal{A}(\Cantor)$, we find that $x \notin \delta_X(A)$ iff $\delta^{-1}(\{x\}) \subseteq A^C$ and choose $\delta_\mathbf{X}$ to be effectively fiber-compact to see that $\delta_\mathbf{X}(A_i) \in \mathcal{A}(\mathbf{X})$ holds uniformly, which concludes the proof.
\end{proof}
\end{proposition}

\begin{proposition}[\cite{debrecht4}]
\label{prop:sigmajump}
For a quasi-Polish space $\mathbf{X}$, the elements of $\mathcal{O}^{(\alpha)}(\mathbf{X})$ are precisely the $\Sigma_{\alpha + 1}^0$-sets in Selivanov's definition.
\end{proposition}

To substantiate the claim that Corollary \ref{corr:syntlhbt} matches the Lebesgue--Hausdorff--Banach Theorem \ref{theo:lhbt}, one further observation is required. The proof of the second part of the statement relies on \cite[Section 4]{brattka}.
\begin{proposition}
Let $\mathbf{Y}$ be computable metric spaces. Then $\operatorname{pw-lim} : \subseteq \mathcal{C}(\mathbb{N}, \mathcal{C}(\mathbf{X}, \mathbf{Y}^{(\alpha)})) \to \mathcal{C}(\mathbf{X}, \mathbf{Y}^{(\alpha+1)})$ is computable. If $\alpha > 1$ and $\mathbf{X}$ is a computable metric space, too, then it admits a computable multi-valued inverse.
\end{proposition}

\subsection{$\Delta_2$-measurability}
Define $\Delta : \subseteq \Baire \to \Baire$ via $\Delta(p)(n) = p(n + 1 + \max \{i \mid p(i) = 0\}) - 1$. Let the finite mindchange endofunctor be defined via $(X, \delta_X)^\nabla = (X, \delta_X \circ \Delta)$ and $(f : \mathbf{X} \to \mathbf{Y})^\nabla = f : \mathbf{X}^\nabla \to \mathbf{Y}^\nabla$. Note that $\mathbb{S}^\nabla = \mathbf{2}^\nabla$, hence the map $^C : \mathcal{O}^\nabla(\mathbf{X}) \to \mathcal{O}^\nabla(\mathbf{X})$ is computable (thus $\mathcal{O}^\nabla(\mathbf{X}) \cong \mathcal{A}^\nabla(\mathbf{X})$. Moreover, $x \mapsto (x, \neg x) : \mathbb{S}^\nabla \to \mathbb{S}' \times \mathbb{S}'$ is computable and computably invertible, which implies that $\mathcal{O}^\nabla(\mathbf{X})$ contains exactly those sets that are both themselves and their complements members of $\mathcal{O}'(\mathbf{X})$, i.e.~corresponds to the $\Delta_2$-sets via Proposition \ref{prop:sigmajump}. Note that $^{\nabla\nabla} \cong ^\nabla$, hence iteration of this endofunctor makes little sense. For computable metric spaces, $^\nabla$-continuity is piecewise continuity as shown in \cite{debrecht5}.

In this context, also separation principles play a r\^ole. Note that $^\nabla$-Hausdorff separation is a uniform counterpart of the $T_D$ separation principle \cite{aull, debrecht2}, it requires that $x \mapsto \{x\} : \mathbf{X} \to \mathcal{A}^\nabla(\mathbf{X})$ is computable. We required one more concept, namely:
\begin{definition}
We call a space $\mathbf{X}$ completely compact, iff it has a total representation $\delta_\mathbf{X} : \Cantor \to \mathbf{X}$.
\end{definition}

Note that any completely compact space is compact (inherited from $\Cantor$), whereas there are compact but not completely compact spaces\footnote{A somewhat trivial example would be a space without computable points.} . In \cite{paulydebrecht}, the following was obtained:
\begin{theorem}[Synthetic Jayne Rogers Theorem]
Any admissible completely compact $^\nabla$-Hausdorff space is $^\nabla$-admissible.
\end{theorem}

\subsection{Markov $\Delta_2^0$-measurability and lowness}
\label{subsec:markovdelta2}
Our next example both shows the need for the concept of a  Markov-computable endofunctor, (as they are not computable endofunctors) and illuminate the r\^ole of adjointness introduced in Subsection \ref{subsec:adjointness}. Let $J : \Cantor \to \Cantor$ be the Turing-jump (i.e. $J(p)$ is the Halting problem relative to $p$), and then define $\int$ via $\int(X, \delta_X) = (X, \delta_X \circ J^{-1})$ with the straight-forward extension to morphisms. This yields a Markov-computable endofunctor.

As observed in more general terms in Subsection \ref{subsec:adjointness}, the computably open subsets of $\int \mathbf{X}$ are just the computably $\Sigma^0_2$-subsets of $\mathbf{X}$. Under this perspective, the space $\int \Cantor$ had already been investigated in \cite{miller}.

Note that $J^{-1}$ is computable, whereas $J$ is not, hence $\id : \int \mathbf{X} \to \mathbf{X}$ is computable, and $\id : \mathbf{X} \to \int \mathbf{X}$ typically not.

Now the low-endofunctor $^\vee$ is defined via $\mathbf{X}^\vee = (\int \mathbf{X})'$. Both $\int$ and $\vee$ were studied in \cite{paulybrattka,brattka4}. The results there are essentially special cases of Proposition \ref{prop:adjoint}: In particular, we see that $(^\vee)(^\vee) \cong ^\vee$, and that if $f$ is $'$-continuous and $g$ is $^\vee$-continuous, then $f \circ g$ is $'$-continuous again (hence the name \emph{low}).

Given that $\id : \int \mathbb{S} \to \mathbf{2}$ and $\id : \{J(0^\mathbb{N}), J(10^\mathbb{N})\} \to \int \mathbb{S}$ are computable; and that $\{J(0^\mathbb{N}), J(10^\mathbb{N})\}' = \mathbf{2}'$,  we find $\mathbb{S}^\vee = (\int \mathbb{S})' = \mathbf{2}' = \mathbf{2}^\nabla = \mathbb{S}^\nabla$. Hence, the \emph{low-open} sets are just the $\Delta_2$-sets again. Thus the  result (originally from \cite{paulydebrecht}) that the Markov $\Delta_2$-measurable functions are the low-computable ones can now be phrased as:

\begin{theorem}
$\Baire$ is Markov $^\vee$-admissible.
\end{theorem}

As separating $^\vee$-continuity and $^\nabla$ continuity on Baire space is straight-forward (it follows from the existence of a low uncomputable sequence), we also see that Markov $d$-admissibility and $d$-admissibility are clearly distinct concepts: $\Baire$ is Markov $^\vee$-admissible and $^\nabla$-admissible, but not Markov $^\nabla$ admissible (and $^\vee$-admissibility is not even defined).

\subsection{Borel equivalence relations as $^{(\alpha)}$-Hausdorff spaces}
Not only are the derived spaces of our theory such as $\mathcal{C}^{'}(\mathbf{X}, \mathbf{Y})$ not admissible, but there are well-studied examples in descriptive set theory of represented spaces that are not admissible (i.e. not understandable as topological spaces). Borel equivalence relations (see e.g. \cite{kechris}) can be defined in our framework as follows:
\begin{definition}
We call a space $\mathbf{X}$ a \emph{Borel equivalence relation}, if it has a total representation and is $^{(\alpha)}$-Hausdorff for some $\alpha$.
\end{definition}

Spaces with total representations have been studied by \name{Selivanov} in \cite{selivanov5}, and spaces that are admissible and $^{(\alpha)}$-Hausdorff by \name{Schr\"oder} and \name{Selivanov} in \cite{selivanov4}. However, a particular Borel equivalence relation of crucial interest is $E_0$ given by $\delta_{E_0}(p) = \delta_{E_0}(q)$ iff $\exists n \ p_{\geq n} = q_{\geq n}$; and $E_0$ is easily seen not to be admissible.

\begin{theorem}[\name{Harrington}, \name{Kechris} \& \name{Louveau} \cite{harrington}]
Let $\mathbf{X}$ be a Borel equivalence relation. Then exactly one of the following holds:
\begin{enumerate}
\item $\exists \alpha \exists f : \mathbf{X} \to \left ( \Cantor \right )^{(\alpha)}$ such that $f$ is a continuous injection.
\item There is a continuous embedding $E_0 \hookrightarrow \mathbf{X}$.
\end{enumerate}
\end{theorem}

A curious phenomenon easily demonstrated on $E_0$ is that $\mathcal{O}(E_0)$ is trivial (i.e. $\{\emptyset, E_0\}$), but for $\alpha > 0$, we find $\mathcal{O}^{(\alpha)}(E_0)$ to carry a Borel-like structure. This shows that descriptive set theory can make sense on represented spaces that have the indiscrete topology as their associated topology, hence are not susceptible to any approach building $\Sigma_2^0$-sets from $\Sigma_1^0$-sets.

\subsection{The endofunctor $\mathcal{K}$}
\label{subsec:compacts}
We shall provide an example of a computable endofunctor $d$ that does change the underlying sets, and consider to what extent notions such as $b$-measurable sets or $d$-continuous functions still make sense. Our example is a very familiar one, namely the operation that takes a represented spaces $\mathbf{X}$ to the space of compact subsets $\mathcal{K}(\mathbf{X})$, and a continuous function $f: \mathbf{X} \to \mathbf{Y}$ to its lifted version. For details, see \cite{pauly-synthetic-arxiv}.

To understand the $\mathcal{K}$-open sets, we need to have a look at $\mathcal{K}(\mathbf{S})$. This space has three elements, $\{\emptyset, \{\top\}, \mathbf{S}\}$, and carries the generalized \Sierp topology, i.e.~$\mathcal{O}(\mathcal{K}(\mathbf{S}))$ has the underlying set $\{\emptyset, \{\emptyset\}, \{\emptyset, \{\top\}\}, \{\emptyset, \{\top\}, \mathbf{S}\}\}$. It seems sensible to interpret $\mathcal{K}(\mathbf{S})$ as a three-valued logic, with $\emptyset$ being \emph{unknown}, $\{\top\}$ being \emph{plausible} and $\mathbf{S}$ being \emph{true}. Thus, a $\mathcal{K}$-open set actually is two open sets, with one contained in the other. Elements of the inner open set are definitely in the $\mathcal{K}$-open, for elements of the outer but not the inner, it is plausible but unknown that they are members of the $\mathcal{K}$-open. Operations such as preimage under continuous functions make sense for such a structure, whereas intersection could be defined in a few different ways -- and as $\mathcal{K}$ does not preserve binary products, we only make claims for the former.

Now let us consider $\mathcal{K}$-continuous functions. A $\mathcal{K}$-continuous function from $\mathbf{X}$ to $\mathbf{Y}$ maps points in $\mathbf{X}$ to compact sets in $\mathbf{Y}$ -- and such mappings have been studied extensively as the upper hemicontinuous maps from $\mathbf{X}$ to $\mathbf{Y}$. Likewise, we may consider the computable endofunctor $\mathcal{V}$ that maps a space to the space of its overt subsets and lifts functions, and would obtain the lower hemicontinuous maps as the $\mathcal{V}$-continuous ones. While our framework does not have many implications for these classes (mainly closure under composition with continuous functions), their example nevertheless indicates that it is unnecessary to restrict our framework to those endofunctors leaving the underlying sets intact (which would be quite problematic for the synthetic part).

\subsection{The analytic sets}
There are various characterizations of the analytic sets in classical descriptive set theory, two of which are particularly relevant for our interests. They can be introduced either as the images of $\Baire$ under a continuous function, or as the projections of closed subsets of $\Baire \times \mathbf{X}$ to the second component. In an effective setting, these two split -- a situation reminiscent of (and ultimately related to) the split of the classical concept \emph{closed set} into \emph{closed set} and \emph{overt set}\footnote{Where we tacitly understand an overt set to be closed in order to uniquely identify it.} in synthetic topology.

First, we shall see that the overt sets, rather than the closed set, occur as the images of $\Baire$ under continuous functions. Unfortunately, we can only prove this for computable metric spaces for now. As shown in \cite{gherardi3}, the identity $\id : \mathcal{A}(\mathbf{X}) \to \mathcal{V}(\mathbf{X})$ is never computable for a non-empty space $\mathbf{X}$, together with the following result, this implies that the first definition of analytic sets cannot yield an extension of the closed sets.

\begin{proposition}[(\footnote{This result is essentially present in \cite{presser}.})]
\label{prop:trace}
The map $\operatorname{Image} : \mathcal{C}(\Baire, \mathbf{X}) \to \mathcal{V}(\mathbf{X})$ is computable and has a computably multivalued inverse with domain $\mathcal{V}(\mathbf{X}) \setminus \{\emptyset\}$ for any complete computable metric space $\mathbf{X}$.
\begin{proof}
That $\operatorname{Image}$ is computable holds true for all represented spaces as a corollary of \cite[Proposition 7.4 (7)]{pauly-synthetic-arxiv}.

For the computability of the inverse, let us be given a non-empty overt set $A \in \mathcal{V}(\mathbf{X})$. Further let $(a_n)_{n \in \mathbb{N}}$ be a computable dense sequence in $\mathbf{X}$. We describe a function $f \in \operatorname{Image}^{-1}(A)$ in terms of a labeled complete countably-branching tree. At the top level, test simultaneously for all $n \in \mathbb{N}$ if $A \cap B(a_n,1) \neq \emptyset$. There is at least one such $n$. Thus, we can obtain an infinite sequence $(n_i)_{i \in \mathbb{N}}$ such that $\{n_i \mid i \in \mathbb{N}\} = \{n \mid A \cap B(a_n,1) \neq \emptyset\}$. We then label the $i$-th child of the root with $B(a_{n_i},1)$. For all subsequent vertices, if the current vertex with depth $k$ is labeled by $B$, we proceed as above, but test $A \cap B \cap B(a_n,2^{-k})$ instead.

From this labeled tree we can find the function $f$ by mapping a path to the unique point in the intersection of all its labels. This is a computable operation due to the properties of complete computable metric spaces, and by construction we find that $f[\Baire] = A$.
\end{proof}
\end{proposition}

The definition of analytic sets as projections however works nicely in our context. We will start by introducing a variant of the \Sierp-space suitable for capturing this class.
\begin{definition}
Let the space $a\mathbb{S} = (\{\bot, \top\}, \delta_{a\mathbb{S}})$ be defined via $\delta_{a\mathbb{S}}(p) = \top$ iff $p$ codes an ill-founded tree, and $\delta_{a\mathbb{S}}(p) = \bot$ otherwise.
\end{definition}

We can extend this definition to yield an endofunctor $a$ making $a\mathbb{S}$ understood as $a$ applied to $\mathbb{S}$ equivalent to $a\mathbb{S}$ defined explicitly above by understanding $a\mathbf{X}$ to be the suitable subspace of $\mathcal{C}(\mathcal{C}(\mathbf{X},\mathbb{S}), a\mathbb{S})$.

\begin{proposition}
For any represented space $\mathbf{X}$ the map $\pi_2 : \mathcal{A}(\Baire \times \mathbf{X}) \to \mathcal{O}^a(\mathbf{X})$ is computable and has a computable inverse.
\end{proposition}

As shown in \cite[Section 5]{pauly-gregoriades-arxiv}, the continuity structure on $\mathcal{O}^a(\mathbf{X})$ for Polish spaces $\mathbf{X}$ corresponds to the structure induced by good universal system as employed in \cite{moschovakis,gregoriades}.

Just like we introduced a Markov-computably-left adjoint Markov-computable endofunctor for $'$ in Subsection \ref{subsec:markovdelta2}, we can introduce a Markov-computably-left adjoint Markov-computable endofunctor for $a$. Pick some standard enumeration $(A_n)_{n \in \mathbb{N}}$ of the computable $\Sigma_1^1$-subsets of Baire, and then let $j_{\textrm{GH}} :\subseteq \Baire \to \Baire$ be defined via $j_{\textrm{GH}}(p) = q$ iff $\{n \mid \exists i \ p(i) = n+1\} = \{n \mid q \in A_n\}$. The map $j_{\textrm{GH}}$ is surjective, and can thus be understood to be a representation -- in fact, it is an admissible representation of the Gandy-Harrington space. A name for a point is an enumeration of all effectively analytic sets containing it.

Next, we introduce $\int_{\textrm{GH}}$ via $\int_{\textrm{GH}}(X,\delta_X) = (X,\delta_X \circ j_{\textrm{GH}})$ and the straight-forward extension to morphisms, and find $\int_{\textrm{GH}}$ to be a Markov-computable endofunctor which is Markov-computably-left-adjoint to $a$. This endofunctor turns effectively $\Sigma^1_1$-sets into effectively open sets -- it seems reasonable to suspect a connection with \name{Gregoriades}' work on turning Borel sets into clopens \cite{gregoriades2}.

\subsection{$K_\sigma$-property and $'$-overtness}
While overtness is a classically invisible condition, its lift along the $'$-endofunctor does yield a topological property similar to the $K_\sigma$-property. For computable Polish spaces, it is in fact equivalent:

\begin{theorem}
A Polish space is $'$-overt iff it is $K_\sigma$.
\begin{proof}
By Corollary \ref{corr:ksigmaimplies'overt} and Lemma \ref{lemma:'overtimpliesksigma}.
\end{proof}
\end{theorem}

\begin{lemma}
\label{lemma:compactimplies'overt}
Let $\mathbf{X}$ be a computably compact computable metric space. Then $\mathbf{X}$ is $'$-overt.
\begin{proof}
Using Proposition \ref{prop:sigma2decomposition}, we can effectively express any $'$-open set $U$ as a union $\bigcup_{n \in \mathbb{N}} A_n$ of closed sets. As $\mathbf{X}$ is computably compact, these closed sets are uniformly compact. By definition of compactness, $\operatorname{IsEmpty} : \mathcal{K}(\mathbf{X}) \to \mathbb{S}$ is computable. Then $\operatorname{IsNonEmpty} : \mathcal{K}(\mathbf{X}) \to \mathbb{S}'$ is computable, too. As $'$ preserves products, we see that $\bigvee : \mathcal{C}(\mathbb{N}, \mathbb{S}') \to \mathbb{S}'$ is computable, and its application yields the final answer.
\end{proof}
\end{lemma}

\begin{corollary}
\label{corr:ksigmaimplies'overt}
Let $\mathbf{X}$ be an effectively $K_\sigma$ metric space. Then $\mathbf{X}$ is $'$-overt.
\begin{proof}
Use Lemma \ref{lemma:compactimplies'overt} together with Proposition \ref{prop:overtunion}.
\end{proof}
\end{corollary}

\begin{lemma}
\label{lemma:'overtimpliesksigma}
Let a Polish space $\mathbf{X}$ be $'$-overt. Then $\mathbf{X}$ is $K_\sigma$.
\begin{proof}
Assume that $\mathbf{X}$ is not $K_\sigma$ and $'$-overt. Then by \cite[Theorem 7.10]{kechris} there is an embedding $\iota : \Baire \to \mathbf{X}$ such that $\iota[\Baire]$ is closed in $\mathbf{X}$. Thus, we may understand $\mathcal{A}(\Baire) \subseteq \mathcal{A}(\mathbf{X}) \subseteq \mathcal{O}'(\mathbf{X})$ (maybe by employing some oracle). That $\mathbf{X}$ is $'$-overt now yields that $\operatorname{IsNonEmpty} : \mathcal{A}(\Baire) \to \mathbb{S}'$ is continuous. Now the names of non-empty closed subsets of $\Baire$ are essentially the ill-founded countably branching trees, which is known to be $\Pi_1^1$-complete (e.g.~\cite[Section 11.8]{bruckner}). But $\operatorname{IsNonEmpty} : \mathcal{A}(\Baire) \to \mathbb{S}'$ being continuous would imply this set to be Borel, contradiction.
\end{proof}
\end{lemma}
\section{Concluding remarks}
Hopefully, our sketch of synthetic descriptive set theory convincingly outlines an exciting new paradigm -- there clearly is much more to do before it can be said to rival classic or effective descriptive set theory when it comes to the breath of the picture painted. A next step would be to obtain a better understanding of the interplay of various computable endofunctors. For example, as the $\Delta_2^0$-sets are just those $\Sigma_2^0$-sets with $\Sigma_2^0$-complement, it seems reasonable to expect a generic way of obtaining $^\nabla$ from $'$. Likewise, it seems desirable to obtain the projective hierarchy from the Borel hierarchy, in particular a synthetic Suslin theorem, as \name{Gregoriades} pointed out.

We should not shy away from pointing out that there is an obstacle to synthetic results implying their classical counterparts, namely uniformity in function measurability: Traditionally, a function $f$ is $\mathfrak{B}$ measurable, if any preimage of an open set is in $\mathfrak{B}$ -- no requirements are imposed on the associated preimage map $f^{-1}$ besides (classically) existing. In a synthetic (or a constructivist) framework, however, such conditions do not appear: The preimage map needs to be a morphism of the underlying category. As discussed already in \cite{paulydebrecht}, when working in the category of represented spaces it is sometimes possible to prove that the classical existence implies continuity, but these proofs can be non-trivial.

\hide{
\section{Biadmissibility}
In order to deal with concepts such as $(\Sigma_n, \Sigma_m)$-measurability, the notion of admissibility needs to be extended further. Now two computable endofunctors appear, $c$ and $d$. We call a function $f$ to be $(c, d)$-measurable, iff the preimages of a $c$-open is uniformly $d$-open.

\begin{proposition}
The canonic map $c\mathbf{X} \to \mathcal{C}(\mathcal{C}(\mathbf{X}, d\mathbb{S}), cd\mathbb{S})$ is computable.
\end{proposition}

\begin{definition}
We call $\mathbf{X}$ $(c, d)$-biadmissible, iff the canonic map $c\mathbf{X} \to \mathcal{C}(\mathcal{C}(\mathbf{X}, d\mathbb{S}), cd\mathbb{S})$ is computably invertible.
\end{definition}

\begin{theorem}
If $\mathbf{Y}$ is $(c, d)$-biadmissible, then for functions from $\mathbf{X}$ to $\mathbf{Y}$ the notions of $(d, cd)$-measurability and $c$-continuity coincide.
\end{theorem}

\begin{theorem}[Jayne Rogers, Second formulation]
Any computable metric space is $( ^\nabla, ')$-biadmissible.
\end{theorem}

As in the definition of $(c, d)$-biadmissibility the left hand side does only depend on $c$, we obtain that $c$ is in a certain sense maximal when applied to $(c, d)$-biadmissible spaces:
\begin{proposition}
Let $b, c, d$ be computable endofunctors such that $cd \cong bd$. Then for any $(c, d)$-biadmissible space $\mathbf{X}$ there is a canonic computable translation $b\mathbf{X} \to c\mathbf{X}$.
\end{proposition}

\begin{corollary}
Let $b$ be a computable endofunctor such that $b\mathbf{X}' \cong \mathbf{X}'$ for any space $\mathbf{X}$. Then for any computable metric space $\mathbf{M}$, we have $b\mathbf{M} \preceq \mathbf{M}^\nabla$.
\end{corollary}

\subsection{Padded representations and the piecewise operator}
Consider the partial map $\operatorname{unpad} : \subseteq \Baire \to \Baire$ defined inductively via $\operatorname{unpad}(0p) = \operatorname{unpad}(p)$ and $\operatorname{unpad}((n+1)p) = n\operatorname{unpad}(p)$. Note that any representation $\delta$ is computably equivalent to $\delta \circ \operatorname{unpad}$, and that the representation $\delta \circ \operatorname{unpad}$ of $\mathbf{Y}$ has the property that any computable realizer of $f : \mathbf{X} \to \mathbf{Y}$ can be extended to a total realizer w.r.t. $\delta \circ \operatorname{unpad}$. We call a representation of the form $\delta \circ \operatorname{unpad}$ \emph{padded}.

Now, given a computable endofunctor $d$ we define another operator $\chi^d$ in the following way. Given a padded representation $\delta$ of a represented space $\mathbf{X}$, the space $\chi^d\mathbf{X}$ uses a pair $(A, (p_i)_{i \in \mathbb{N}}) \in \mathcal{A}^d(\mathbb{N}) \times \mathcal{C}(\mathbb{N}, \Baire)$ as a name for $x \in X$ when $A \neq \emptyset$ and $\forall i \in A \ \delta(p_i) = x$. (\emph{I am assuming here that the notion of element makes sense for $\mathcal{A}^d(\mathbb{N})$, this is ensured by $\{0, 1\}$ being the underlying set of $d\mathbb{S}$.})

The operator $\chi^d$ corresponds to the concept of \emph{piecewise}, in particular we have $\chi^{\id} = ^\nabla$ as a consequence of the fact that $\C_\mathbb{N}$ is complete for finite mindchange computability. More generally, we can simply compose $\chi^d$ with some other operator $c$ to obtain higher level versions: Using $\chi^{(n)}(\mathbf{X}^{(m)})$ on the codomain should results in $\Pi_{n + 1}^0$-piecewise $\Sigma_m^0$-measurability.

\begin{conjecture}
If $\beta < \alpha$, then $\chi^{(\beta)}\mathbf{X}^{(\alpha)} \cong \mathbf{X}^{(\alpha)}$.
\end{conjecture}

\begin{proposition}
\label{piecewiseequality}
Let $\wedge : d\mathbb{S} \times d\mathbb{S} \to d\mathbb{S}$ and $\bigvee : \mathcal{C}(\mathbb{N}, d\mathbb{S}) \to d\mathbb{S}$ be computable, and let $d\mathbb{S}$ have a total padded representation. If $\neq : \mathbf{X} \times \mathbf{X} \to d \mathbb{S}$ is computable, then $\neq : \mathbf{X} \times \chi^d\mathbf{X} \to c\mathbb{S}$ is computable.
\begin{proof}
We can apply the assumed equality test to the first argument paired with each entry in the list partially constituting a $\chi^d\mathbf{X}$-name of the second argument. As $d\mathbb{S}$ has a total padded representation, we do not need to worry whether the entries actually encode elements in $\mathbf{X}$ or not. Thus, we obtain a list of answers in $\mathcal{C}(\mathbb{N}, d\mathbb{S})$, by type conversion we obtain the set in $\mathcal{A}^d(\mathbb{N})$ of indices equal to the first argument. Now we can compute the intersection with the set of correct indices which is part of the $\chi^d\mathbf{X}$-name. Applying $\bigvee$ to this set provides the intended answer.
\end{proof}
\end{proposition}

\begin{theorem}
Let $d\mathbb{S}$ have a total padded representation, $\wedge : d\mathbb{S} \times d\mathbb{S} \to d\mathbb{S}$ and $\bigvee : \mathcal{C}(\mathbb{N}, d\mathbb{S}) \to d\mathbb{S}$ be computable. Let $c\mathbf{Y}$ have a padded representation $\delta_{c\mathbf{Y}}$ such that $\dom(\delta_{c\mathbf{Y}}) \in \mathcal{A}^d(\Baire)$ is computable. Let $\neq : c\mathbf{Y} \times c\mathbf{Y} \to d \mathbb{S}$ be computable. Then the following two conditions are equivalent:
\begin{enumerate}
\item $f : \mathbf{X} \to \chi^dc\mathbf{Y}$ is computable.
\item $f : \mathbf{X} \to c\mathbf{Y}$ is non-uniformly computable, and $y \mapsto f^{-1}(\{y\}) : c\mathbf{Y} \to \mathcal{A}^{d}(\mathbf{X})$ is computable.
\end{enumerate}
\begin{proof}
\begin{description}
\item[$1. \Rightarrow 2.$] A $(\mathbf{X}, \chi^dc\mathbf{Y})$-realizer of $f$ produces (among other things) a list of elements, one of which is a name for $f(x) \in c\mathbf{Y}$. Hence, $f : \mathbf{X} \to c\mathbf{Y}$ has to be non-uniformly computable.

    To see that $y \mapsto f^{-1}(\{y\}) : c\mathbf{Y} \to \mathcal{A}^{d}(\mathbf{X})$ is computable, we can alternatively show that $(x, y) \mapsto (f(x) \neq y) : \mathbf{X} \times c\mathbf{Y} \to d\mathbb{S}$ is computable. Function application with the assumption, together with $\neq : \chi^dc\mathbf{Y} \times c\mathbf{Y} \to d\mathbb{S}$ obtained from Proposition \ref{piecewiseequality} does the job.
\item[$2. \Rightarrow 1.$] If we consider $c\mathbf{Y}$ to be equipped with a padded representation, we can use a name of $x \in \mathbf{X}$ to compute a sequence $(p_i)_{i \in \mathbb{N}}$ of sequences, such that at least one entry is a name for $f(x) \in c\mathbf{Y}$, since $f$ is non-uniformly computable (just try any code for a Turing machine, and extend the partial functions to total ones). This sequence constitutes the second part of a name for $f(x) \in \chi^dc\mathbf{Y}$.

    The computation described in the following results in some element of $d\mathbb{S}$. As this space has a total padded representation by assumption, we can ensure to obtain a well-formed end result, even if intermediate values are outside the domains of the computable functions applied to them. Starting from $i \in \mathbb{N}$, we consider $y_i = \delta_{c\mathbf{Y}}(p_i)$, and use the assumption to compute $f^{-1}(\{y_i\}) \in \mathcal{A}^d(\mathbf{X})$. Then we may test $x \notin f^{-1}(\{y_i\})$ (i.e. $f(x) \neq y_i$) and obtain the answer as an element of $d\mathbb{S}$. Simultaneously we test $y_i \notin \dom(\delta_{c\mathbf{Y}})$, again obtaining the answer in $d\mathbb{S}$. Apply $\wedge$ on the latter two results to obtain $b_i \in d\mathbb{S}$.

    The computation above produces a function $A : \mathbb{N} \to d\mathbb{S}$, which we interpret as $A \in \mathcal{A}^d(\mathbb{N})$. This set constitutes the first part of the desired name for $f(x) \in \chi^dc\mathbf{Y}$.
\end{description}
\end{proof}
\end{theorem}

\begin{observation}
Let $c$ satisfy that $\id : \mathbf{Y} \times c\mathbf{Y} \to c(\mathbf{Y} \times \mathbf{Y})$ is computable. If $\mathbf{Y}$ is computably Hausdorff, then $y \mapsto \{y\} : c\mathbf{Y} \to \mathcal{A}^c(\mathbf{Y})$ is computable.
\end{observation}

\begin{observation}
Let $c$ satisfy that $\id : \mathbf{Y} \times c\mathbf{Y} \to c(\mathbf{Y} \times \mathbf{Y})$ is computable and $\mathbf{Y}$ be computably Hausdorff. If $f$ is computably $(c, d)$-measurable, then $y \mapsto f^{-1}(\{y\}) : c\mathbf{Y} \to \mathcal{A}^{d}(\mathbf{X})$ is computable.
\end{observation}}
\bibliographystyle{eptcs}
\bibliography{../spieltheorie}
\section*{Acknowledgements}
The first author would like to thank Vasco Brattka, Vassilis Gregoriades, Takayuki Kihara, Luca Motto-Ros, Matthias Schr\"oder and Victor Selivanov for fruitful dicussions on the subject of the present paper. The work has benefited from the Marie Curie International Research Staff Exchange Scheme \emph{Computable
Analysis}, PIRSES-GA-2011- 294962.
\end{document}